\documentclass[%
 reprint,
groupedaddress,
showpacs,
 amsmath,amssymb,
 aps,
prb,
]{revtex4-1}

\usepackage{graphicx}
\usepackage{dcolumn}
\usepackage{bm}

\begin{document}

\newcommand{\be}{\begin{equation}}
\newcommand{\ee}[1]{\label{#1}\end{equation}}
\newcommand{\bem}{\begin{eqnarray}}
\newcommand{\eem}[1]{\label{#1}\end{eqnarray}}
\newcommand{\eq}[1]{Eq.~(\ref{#1})}
\newcommand{\Eq}[1]{Equation~(\ref{#1})}
\newcommand{\vp}[2]{[\mathbf{#1} \times \mathbf{#2}]}


\title{Transverse forces on a vortex in lattice models of superfluids}

\author{E.  B. Sonin}
 \affiliation{Racah Institute of Physics, Hebrew University of
Jerusalem, Givat Ram, Jerusalem 91904, Israel}

\date{\today}

\begin{abstract}
The paper derives the transverse forces (the Magnus and the Lorentz forces) in the lattice models of superfluids in the continuous approximation.  The continuous approximation  restores translational invariance absent in the original lattice model, but the theory is not Galilean invariant. As a result, calculation of the two transverse forces on the vortex, Magnus force and Lorentz force,  requires the analysis of two balances, for the  true momentum of particles in the lattice (Magnus force) and for the quasimomentum (Lorentz force) known from the Bloch theory of particles in the periodic potential. While the developed theory yields the same Lorentz force, which was well known before,  a new general expression for the Magnus  force was obtained. The theory demonstrates how a small Magnus force emerges in the Josephson-junction array if the particle-hole symmetry is broken. The continuous approximation for the Bose--Hubbard model close to the superfluid-insulator transition was developed, which was used for calculation of the Magnus force.  The theory shows that there is an area in the phase diagram  for the Bose--Hubbard model, where the Magnus force has an inverse sign with respect to that which is expected from the sign of velocity circulation.   \end{abstract}

\maketitle


\section{Introduction} \label{Intr}

The transverse force on a vortex in superfluids (neutral and charged)  is debated during many decades and has been a topic of reviews and books \cite{RMP,Kop,Magn,VolB}. In a continuous superfluid at $T=0$, which is identical to a perfect fluid in classical hydrodynamics, the balance of forces on a vortex is 
\be
\bm F_M + \bm F_L= m n \left[\left(\bm  v_L-\bm v_s  \right)\times \bm  \kappa\right] = \bm F_{ext},
     \ee{B}
where the Magnus force $\bm F_ M$ is proportional to the vortex velocity $\bm v_L$ and the Lorentz force $\bm F_L$ is proportional to the superfluid velocity $\bm v_s ={\hbar \over m} \bm \nabla \varphi$ determined by the phase $\varphi$ of the order parameter wave function, and the external force $ \bm F_{ext}$ combines all other forces on the vortex, e.g., pinning and friction forces. Here $n$ is the density of particles with mass $m$, and $\bm \kappa$ is the vector parallel to the vortex axis with its modulus equal to the circulation quantum $\kappa=h/m$. The united transverse force $\bm F_M + \bm F_L$ depends only on the relative velocity $\bm v_L-\bm v_s$ as required by the Galilean invariance. 

In lattice  models of superfluids the Galilean invariance is absent, and  the value of the Magnus force   was under scrutiny. The most known lattice model of the superfluid is the Josephson junction array. Usually they studied  vortex dynamics in the array in the continuous approximation. These studies have not revealed any Magnus force normal to the vortex velocity \cite{Eck}. Moreover, there was  experimental evidence of the ballistic vortex motion in the Josephson junction array  \cite{2}, which is possible only in the absence of the Magnus force.  In the classical theory of the Josephson junction array they  usually assumed the particle-hole symmetry, which  forbids the Magnus force in the model (see \cite{PRB7} and references therein). However, this symmetry is not exact, and there  was a lot of theoretical works aiming at finding  a finite Magnus force, mostly suggesting some quantum effects. In superconductors the Magnus force determines the Hall effect, and the presence or the absence of this force means the presence or the absence of the Hall effect.

Intensive investigations of Bose-condensed cold atoms attracted an interest to another lattice model of a superfluid:  the Bose--Hubbard model \cite{FishH}. The periodic structure of potential wells for bosons, which leads to the Bose--Hubbard model in the tight-binding limit,  is realized for cold-atom BEC in experiments with optical lattices \cite{Ued}. Recently Lindner {\em et al.} \cite{Auer} and Huber and Lindner \cite{Lind} calculated the Magnus force in the Bose--Hubbard model and revealed that close to the superfluid-insulator transition the force changes its sign as happens in Fermi superfluids at changing the sign of the carrier charge.

The paper presents the analysis of the transverse forces on the vortex in a lattice, which is approximated by a continuous model. The forces are determined from the momentum balance. The absence of the Galilean invariance makes necessary to analyze two momentum balances: for true momentum and for quasimomentum known from the Bloch band theory for particles in a periodic potential. This yields the general expression for the Magnus force in the absence of the Galilean invariance, which was used for calculation of
 the Magnus force in the Bose--Hubbard model close to the superfluid-insulator transition.

\section{Vortex dynamics in the continuous approximation for the lattice superfluid} \label{CML}

The continuous approximation for lattice superfluids generally enough gives the phenomenological theory, which corresponds to the Lagrangian:
\begin{equation}
L=- \hbar n \dot \varphi 
 - {\hbar^2  \tilde n \over 2m}(\bm \nabla \varphi )^2 -E_c(n),
                   \label{contL} \end{equation}
where  $E_c(n)$ is the energy of a resting liquid which depends only on $n$. For simplicity we consider the 2D problem, where $n$ is the particle number per unit area.
The Hamiltonian (energy) for this Lagrangian is
\begin{equation}
H=  {\partial L\over \partial \dot \varphi} \dot \varphi -L={\hbar^2  \tilde n \over 2m}(\bm \nabla \varphi )^2+E_c(n) .
                   \label{contHh} \end{equation}
Despite similarity of the model to hydrodynamics of the perfect fluid, there is an essential difference. The continuous approximation for the lattice model restores translational invariance but not Galilean invariance. The latter is absent since the effective density $\tilde n$, which characterizes stiffness of the phase field,  is different from the true particle density $n$. This difference is an attribute of any lattice model, and  the effective density $\tilde n$ is much less than $n$ if the lattice nodes are weakly connected.

 Let us discuss the conservation laws, which follow from  Noether's theorem. 
  The gauge invariance provides the conservation law for charge (particle number):
\be
{\partial \over \partial t}{\partial L\over \partial \dot \varphi} +\nabla_k\left({\partial L\over \partial \nabla_k \varphi} \right)=0. 
    \ee{}
This  is the continuity equation (the first Hamilton equation) for the fluid:
\be
m{\partial n \over \partial  t}=-\bm \nabla \cdot {\bm j} .
    \ee{nt}
where $n$ is the particle density and
\be
\bm j=-{m\over \hbar}{\partial L\over \partial \bm\nabla \varphi} =\hbar \tilde n\bm \nabla \varphi
   \ee{Gcur}
is the mass current. The mass current, which by the factor $m/q$ differs from the charge current of particles with the charge $q$, is at the same time the momentum density.  

The second Hamilton equation for the phase $\varphi$ canonically conjugate to $n$ is
\be
\hbar {\partial \varphi \over \partial t}= -{\partial H\over \partial n }=-\mu- {\hbar^2\over 2m }{d\tilde n\over dn}  (\bm \nabla \varphi)^2,
         \ee{nph}
where $\mu =\partial E_c(n)/\partial n$ is the chemical potential of the liquid at rest.

The translational invariance provides the conservation law
\bem
{\partial g _k \over \partial t} +\nabla_l\Pi_{kl}=0,
     \eem{NT}
 for the momentum with the density (current)
\be 
\bm g=-{\partial L\over \partial \dot \varphi}\bm\nabla  \varphi =\hbar n\bm \nabla \varphi.
         \ee{}
Here the momentum-flux tensor is
\bem 
 \Pi_{kl}={\partial L\over \partial \nabla_k\varphi} \nabla_l\varphi-L\delta_{kl}={\hbar^2\over  m  } \tilde n\nabla_k\varphi  \nabla_l\varphi
 \nonumber \\
+\left[P  +{\hbar^2\over 2m }\left({d \tilde n\over dn} n-\tilde n\right)(\bm \nabla \varphi)^2  \right] \delta_{kl},
   \eem{MF}
and the pressure $P$ is connected with the chemical potential $\mu$ by the $T=0$ thermodynamic  Gibbs-Duhem  relation $d P = n d\mu$. 
  
In the Galilean invariant system the current $\bm g$, which appears  in the Noether conservation law following from the translation invariance, coincides with $\bm j$. But in our case with broken Galilean invariance ($\tilde n \neq n$) the currents $\bm g$ and $\bm j$ differ. The true mass current (true momentum density) is $\bm j$ but not $\bm g$. This follows from the fact that  the density $n$ and the current $\bm j$ in the continuity equation (\ref{nt}) are nothing else as  averages of their relevant quantum mechanical operators $\hat n ={\hat \psi}^\dagger \ {\hat \psi}$ and 
\be
\hat {\bm j}=-{i\hbar \over 2}({\hat \psi}^\dagger \bm \nabla {\hat \psi}  - \bm \nabla{\hat \psi}^\dagger {\hat \psi}) , 
   \ee{qmCur}
where ${\hat \psi}$ and ${\hat \psi}^\dagger$ are the annihilation and the creation operators normalized to the density.
The continuity equation (\ref{nt}) is universal and  valid for {\em any} gauge-invariant system including lattice superfluids independently from what forces are applied to the system or how particles interact. 

    So Noether's theorem for a translational invariant but not Galilean invariant system does not provide the conservation law for the true momentum. In the next section we shall see that for particles in a periodic  potential    the current $\bm g$ coincides with the quasimomentum density.  
    
Although   Noether's theorem does not lead to the conservation law for the true momentum, the true  momentum conservation law, nevertheless,  approximately takes place as can be checked using the Hamilton equations (\ref{nt}) and (\ref{nph}) and neglecting higher than second order in phase gradients terms:
\bem
{\partial j _k \over \partial t} +\nabla_l \tilde\Pi_{kl}=0,
     \eem{jk}
where the momentum-flux tensor is
\bem 
\tilde \Pi_{kl}={\hbar^2\over  m  } {d \tilde n\over dn}\tilde n\nabla_k\phi  \nabla_l\varphi
+\tilde P \delta_{kl},
   \eem{MFg}
and the partial pressure $\tilde P $ is determined by the relation $d\tilde P =\tilde n d\mu$. 

The most reliable method to derive the equation of vortex motion is to consider the momentum balance.
The momentum balance requires that any external force on a vortex is compensated by  the momentum flux through a cylindric surface surrounding the vortex line \cite{RMP,PRB7}. The problem with superfluids on lattices is that there is  a momentum exchange between the superfluid and the system, which provides the periodic lattice potential. The continuous approximation, which restores translational invariance, in fact takes into account this momentum exchange since translational invariance leads to the conservation law for the Noether momentum (quasimomentum) but not the true momentum of particles.  
 We argue that the Lorentz and the Magnus force must be derived from the balance of different momenta: the quasimomentum for the former and the true momentum for the latter.  Deriving the Lorentz force one can assume that the vortex is at rest in the laboratory coordinate frame connected with the lattice. Solving \eq{nph} for the time-independent phase $\varphi$ one obtains the quadratic in $\bm \nabla \varphi $ correction to the chemical potential (Bernoulli's  effect):
\be
\mu' =- {\hbar^2\over 2m }{d\tilde n\over dn}  (\bm \nabla \varphi)^2.
         \ee{nph0}
Then the momentum-flux tensor   (\ref{MF}) becomes 
\bem 
 \Pi_{kl}={\hbar^2\over  m  } \tilde n\nabla_k\phi  \nabla_l\varphi
+\left[P_0-{\hbar^2\over 2 m  }\tilde n(\bm\nabla \varphi)^2  \right] \delta_{kl}, 
   \eem{MF1}
where $P_0$ is a constant pressure in the absence of any velocity field. The components of the Lorentz force  are given by the integral over a cylinder around the vortex: $F_{Li} = \oint  \Pi_{kl} dS_l$. The phase gradient  $\bm\nabla  \varphi= \bm \nabla\varphi_v  + \bm \nabla\varphi_t $ consists of the gradient $\bm \nabla\varphi_v =[\hat z \times \bm r]/ r^2$ induced by the vortex line and the gradient $\bm \nabla\varphi_t  =\bm j /\hbar  \tilde n$ produced by the transport current. The force arises from the cross terms $\bm \nabla\varphi_v  \cdot \bm \nabla\varphi_t $ in the momentum flux tensor. Their  integration yields 
\be 
\bm F_L=- [\bm  j  \times \bm  \kappa]=- m\tilde n[\bm  v_s  \times \bm  \kappa].
       \ee{MagnL}
The Lorentz force follows from the quasimomentum balance because it is a momentum  exchange between the transport velocity field and the vortex. But any variation of the transport velocity must be accompanied by the momentum transfer to or from the lattice as follows, e.g., from the Bloch band theory for particles in  a periodic potential (see the next section).

Deriving the Magnus force proportional to the vortex velocity we can consider the case when the superfluid does not move with respect to the lattice. Then it is natural to expect that there is no momentum exchange between the superfluid and the lattice. Therefore one may conclude that the derivation of the Magnus force requires the balance of the true momentum of particles. It is more convenient to consider this balance in the coordinate frame moving with the vortex since only in this frame the state is stationary, at least in average.  The law of the Galilean transformation (see the next section) requires that the expressions for the  energy, \eq{contHh}. and the momentum-flux tensor, \eq{MFg}, remain valid in the coordinate frame moving with the velocity $\bm w$ if the phase gradient $\bm \nabla \varphi$ is replaced by  the phase gradient $\bm \nabla \varphi'=\bm \nabla \varphi - m \bm w /\hbar$  in the moving frame. Following the same steps as at derivation of the Lorentz force in the laboratory frame, one obtains that the momentum transferred to the liquid, which is the Magnus force in our case,  is proportional to the  phase gradient $\bm \nabla \varphi'_t$ connected with the transport supercurrent in the moving frame. Since in the laboratory frame the transport supercurrent is absent ($\bm \nabla \varphi_t=0$) and $\bm w=\bm v_L$, calculating the Magnus force components $F_{Mi} = \oint  \tilde \Pi_{kl} dS_l$ one obtains the Magnus force
\be 
\bm F_M={d\tilde n\over dn} \tilde n m[\bm  v_L  \times \bm  \kappa].
       \ee{Mf}
The force appears  due to convection of the vortex-related momentum $\hbar \tilde n \bm \nabla \varphi_v$   into the area of the momentum balance by the supercurrent. Since the circular velocity field ${\hbar\over m}\nabla \varphi_v$ is fixed, an arrival of a particle into the balance area may change the vortex-related momentum only via variation of $\tilde n$ and does not require accompanying momentum transfer to the lattice. This is another argument why the Magnus force is determined from the balance of the true momentum and is proportional to $d\tilde n/ dn$. 

In the Josephson junction array the current between two nodes of the lattice is determined by the Josephson energy $E_J\cos(\varphi_1-\varphi_2)$, where $\varphi_1$ and $\varphi_2$ are the phases at the two nodes. In   the continuous limit this yields   $ \tilde n =m E_J/\hbar^2$. The particle-hole symmetry requires that $E_J$ and    $ \tilde n $ do not depend on the average density  $n$, and the Magnus force vanishes in agreement with the symmetry of this model. 
 So our approach yields correct values of the Magnus force at least in the two  limits when the  force is known exactly:  the Galilean invariant liquid and the Josephson junction array with particle-hole symmetry. 

Without external forces $\bm F_L +\bm F_M=0$, and the vortex moves with the velocity  by the factor ${d\tilde n/ dn}$ less than the superfluid velocity $\bm v_s$. This shows that Helmholtz's theorem (the vortex moves with the fluid velocity) is not valid without Galilean invariance.

{\section{Vortex dynamics from the Bloch band theory} \label{Bloch}

The meaning of our approach becomes more transparent if one applies it  to particles in a periodic potential $U(\bm r)=U(\bm r+\bm a)$.  Here $\bm a$ can be any of the translation vectors of the periodic structure. The density of the
single-particle energy is 
\be
E_s=  {\hbar^2 \over 2m}|\bm \nabla \psi|^2+U(\bm r) | \psi|^2.
                   \ee{}
The eigenstates are described by Bloch functions: 
\be
\psi(\bm r,t) =u_n(\bm r, \bm  k)e^{i \bm k\cdot \bm r-iE(\bm k) t/\hbar},
           \ee{BF}
where $u_n(\bm r, \bm  k)$ is a periodic function. 
The quasimomentum $\hbar \bm k$ differs from the true momentum of the quantum state. The latter can be calculated averaging, i.e., integrating   the quantum mechanical expression (\ref{qmCur}) for the momentum operator over the crystal unit cell:
\be
\bm p= -i\hbar \int \psi(\bm r)^*\bm \nabla \psi(\bm r)d\bm r = \hbar \bm k   -i\hbar \int u(\bm r)^*\bm \nabla u(\bm r)d\bm r .
      \ee{}
Here, in contrast to \eq{qmCur}, the wave function is normalized to the unit cell,: $\int |\psi|^2 d\bm r=1$. 
Calculating the band energy $E(\bm k)$ in the $\bm k \bm p$ approximation for small $k$, i.e., at the band bottom (the energy minimum at $k=0$)  one obtains that 
\be
E(\bm k)={\hbar^2 k^2\over 2m^*},~~\bm p =m \bm v_g,~~\bm v_g= {d^2 E(\bm k )\over \hbar d\bm k^2 } \bm k = {\hbar \bm k\over m^*},
    \ee{}
 where  $\bm v_g $ is the group velocity  and $m^*$ is the effective mass.  Suppose that particles are bosons, which condense in a single Bloch state with density $n$. The wave vector $\bm k =\bm \nabla \varphi$ is a gradient of the phase $\varphi$.  Then the true momentum density (mass current) $\bm j=n \bm p$  coincides with that given by \eq{Gcur} if  $\tilde n =n m/m^*$.   
On the other hand, the current $\bm g = n\hbar \bm k$ is the quasimomentum density. So the Bloch band theory for bosons clearly connects the currents $\bm j$ and $\bm g$ derived in Sec.~\ref{CML} from the phenomenological Lagrangian with the densities of the true momentum and the quasimomentum respectively.

It is  well known from the solid state physics that an external force on a particle in the energy band determines time variation of  the quasimomentum but not the true momentum:
\be
\hbar {d\bm k\over dt}= m^*{d \bm v_g\over dt} = \bm f.
      \ee{New}
In the absence of Umklapp processes the total quasimomentum is also a conserved quantity, and the conservation law  for the quasimomentum of the Bose-condensate in a single Bloch state is given by \eq{NT}. Only  the part $d\bm p/dt$ of the whole momentum variation $\hbar d\bm k/dt$ brought to the system by the external force is transferred to particles. The rest is transferred to the lattice supporting the periodic potential. It is worthwhile of noticing that in the Bloch theory for particles in a periodic potential the true momentum differs from the quasimomentum by the constant factor $m/m^*$. Therefore the true momentum conservation law is exact and directly follows from the quasimomentum (Noether's) conservation law after multiplying the latter by $m/m^*$. But in the general case considered in the previous section  only the quasimomentum conservation law was exact. 

At the Galilean transformation to the coordinate frame moving with the velocity $\bm w$ ($\bm r=\bm r'+\bm w t$, $t=t'$) the Hamiltonian and the Schr\"odinger equation retain their form, but the wave function must transform as
$ \psi=\psi'e^{im\bm v\cdot \bm r' /\hbar +im v^2 t/2\hbar}$. Correspondingly, in the moving coordinate frame the Bloch  function (\ref{BF}) becomes 
\be
\psi'(\bm r',t') =u_n(\bm r'+\bm w t', \bm  k)e^{i \bm k'\cdot \bm r'-iE_f(\bm k)\hbar},
           \ee{BFt}
where the wave vector $\bm k'$ and the energy $E_f (\bm k)$ are connected with those in the laboratory frame by the relations 
\bem
 \bm k'=\bm k-{m\over \hbar}\bm w,~~E_f=E(\bm k) -\hbar \bm k \cdot \bm w +{mw^2\over 2}
 \nonumber \\
\approx {  (\hbar \bm k-m^*\bm w)^2\over 2m^*}+{(m-m^*)w^2\over 2} .
     \eem{}
In the moving frame the Schr\"odinger equation  contains the time dependent periodic potential, and its solution is not an eigenstate of the quantum mechanical energy operator $i\hbar \partial /\partial t$. \Eq{BFt} is a solution of this equation following from the Floquet theorem, the energy $E_f$ being the Floquet quasienergy.   The  true energy of the state is an average value (expectation value)  of the energy operator   independently from whether the  state is an eigenstate of the energy operator, or not. It is different from the Floquet quasienergy and is given by
\bem
E' = \int \psi'^* i\hbar{\partial \psi \over \partial t}d\bm r= E_f +  \int u_n ^* i\hbar(\bm w \cdot \bm \nabla) u\,d\bm r=E(\bm k)
\nonumber \\
 -\bm p \cdot \bm w +{mw^2\over 2}\approx {\hbar ^2 (\bm k-m\bm w)^2\over 2m^*}+{mw^2\over 2}\left(1-{m\over m^*}\right).
 \nonumber \\
     \eem{enM}
The Galilean transformation demonstrates the difference between the quasimomentum and the quasienergy on one side, and  the true momentum and the true energy on the other. If one ignores the difference and treats the quasiparticle as a real particle with the mass $m^*$ the particle current $\bm j'/m=n (\hbar \bm k/m^*- \bm w) $ in the the ground state in the moving frame vanishes. But at the minimum of the true  energy given by \eq{enM} the particle current  $\bm j'/m$ does not vanish. This is the effect of dragging of particles by a moving periodic potential, which is especially pronounced in the limit of infinite effective mass when the particles are totally trapped by the periodic potential and cannot move relatively to it. The  effect  was observed for a potential produced by  a running  acoustic wave, which drags electrons (acoustoelectric effect) \cite{AEE} or excitons \cite{AcusEx}. 

The whole analysis of this section addressed only single-particle states and the ideal Bose--Einstein condensation in a single-particle state. But a real vortex with a well-defined core is impossible without interaction. However, adding weak particle-particle interaction one can develop the Gross--Pitaevskii theory similar to that theory for uniform translational invariant liquids. This approach is valid as far as the interaction is not too strong and  the vortex-core radius essentially exceeds the lattice period. 

\section{Vortex dynamics in the Bose--Hubbard model}

The Hamiltonian of the  Bose--Hubbard model \cite{FishH} for a lattice with distance $a$ between nodes is
\bem
{\cal H}= -J\sum _{i,j} \hat b_i^\dagger  \hat b_j +{U\over 2}\sum_i \hat N_i(\hat N_i-1)- \mu  \sum_i \hat N_i.
    \eem{BH}
Here $\mu$ is the chemical potential, the operators $\hat b_i$ and $\hat b_i^\dagger$ are the operators of annihilation and creation of a boson at the $i$th lattice node,  and $\hat N_i=\hat b_i^\dagger\hat b_i$ is the particle number operator at the same node. The first sum is over neighboring  lattice nodes   $i$ and $j$.

In the superfluid phase with large numbers of particles $N_i$ all operator fields can be replaced by the classical fields in the spirit of the Bogolyubov theory:
\be   
\hat b_i~\to~\sqrt{N_i}e^{i\varphi_i},~~~\hat b_i^\dagger~\to~\sqrt{N_i}e^{-i\varphi_i},
        \ee{qaN}
where $\varphi_i$ is the phase at the $i$th node. After transition to the continuous approach one obtains the Hamiltonian (\ref{contHh}), where\footnote{We assume that there is the same number of particles at all nodes and write $N_i$ without the subscript $i$.}
\be
n={N\over a^2},~~\tilde n= {mz_0Ja^2 \over \hbar^2}n,~~E_c(n)={Ua^2\over 2}n^2- \mu  n. 
    \ee{}
Here $z_0$ is the number of nearest neighbors equal to 4 in the quadratic lattice. This  is the tight-binding  limit of the Bose condensate of particles in a Bloch state (see the previous section) when the effective mass is 
\be
m^*= {\hbar^2\over z_0Ja^2}. 
    \ee{mass}

\begin{figure}
\includegraphics[width=.5\textwidth]{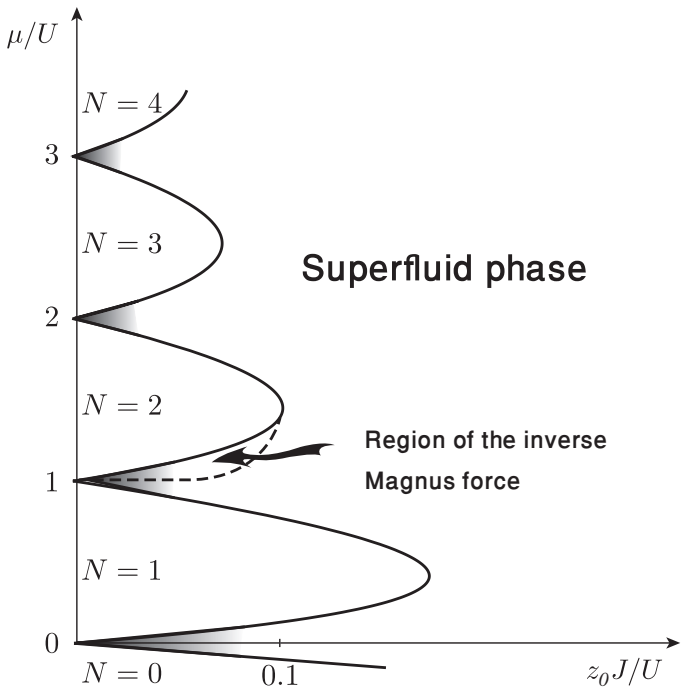}
\caption[]{ The phase diagram of the Bose--Hubbard model.  The Mott insulator phase occupies lobes corresponding to fixed integer numbers $N$ of bosons. The shaded beaks of the superfluid phase, which penetrate between insulator lobes, are analyzed in the text. The dash line separates the region with the inverse Magnus force from the rest of the superfluid phase. The line is schematic since it was really calculated only in the limit $J\to 0$ where it is horizontal. The region of the inverse Magnus force  exists under any lobe but is shown only for the beak between the $N=1$ and  $N=2$ lobes. }
\label{f12-0}
\end{figure}

When the energy $J$ of the internode hopping decreases, the phase transition from superfluid to Mott insulator must occur \cite{FishH}.   In the limit $z_0J/U \to 0$
when the hopping term $\propto J$ can be ignored the eigenstates are given by Fock states $|\Psi_N\rangle= |N\rangle$ with fixed number $N$ of particles at any node. At growing $J$ the transition line can be found in the mean-field approximation \cite{Ued}. One takes into account the  hopping term  introducing the mean field equal to the average value of the annihilation operator (and its complex conjugate the creation operator):
\be   
\langle \hat b_i\rangle=\psi_i =|\psi| e^{i\varphi_i},~~~ \langle \hat b_i^\dagger\rangle=\psi_i^*   =|\psi| e^{-i\varphi_i}.
        \ee{qa}
It is assumed that only the phase but not the modulus of the order parameter $\psi$ varies from  node to  node. In general $|\psi|^2 $ is not equal to $N$ as   \eq{qaN} assumes and must be determined from the the condition of self-consistency (see below).  Introducing the mean field one reduces the problem to the single-node problem with the Hamiltonian
\bem
{\cal H}_s= -zJ(\hat b^\dagger \psi +\psi^*  \hat b)+{U\over 2} \hat N(\hat N-1)- \mu  \hat N.  
    \eem{BHs}
Here 
\be 
z=\sum_j e^{i(\varphi_j-\varphi_i)}
   \ee{}
reduces to the  number $z_0$ of nearest neighbors in the uniform state with the constant phase at all nodes. 

The multi-node wave function is a product of the single-node wave functions. Calculating the energy of the original Hamiltonian (\ref{BH}) for this wave function and minimizing it with respect to $\psi$ one obtains the self-consistency equation, which determines $\psi$. Following this approach \cite{Ued} one obtains the phase diagram  shown in Fig.~\ref{f12-0}. The Mott-insulator phases with fixed numbers $N$ of particles  per node occupy interiors of lobes at small $z_0J/U$. 

We address vortex dynamics close to the phase transition at minimal values of $J$, i.e. at beaks of the superfluid phase between  lobes, which are shaded in  Fig.~\ref{f12-0}. Here the mean-field approximation is simplified by the fact that only two states with $N$ and $N+1$ particles  interplay in the beak between the lobes $N$ and $N+1$. This is because at $\mu=NU$ these two states have the same energy, whereas all  other states are separated by a gap on the order of the high energy $U$.  So for a beak between the lobes corresponding to Mott insulators with the number of bosons per node $N$ and $N+1$ we look for a solution in the form of a superposition of two Fock states:
\be
|\Psi_N\rangle= f_N|N\rangle +f_{N+1}|N+1\rangle.
   \ee{}
This wave function is an eigenfunction of the single-node Hamiltonian (\ref{BHs}) if 
\bem
f_{N,N+1}=\sqrt{ \sqrt{{\mu'^2\over 4}+z^2J^2(N+1) |\psi|^2}\mp {\mu' \over 2}\over {2 \sqrt{{\mu'^2\over 4}+z^2J^2(N+1) |\psi|^2}}}, 
    \eem{}
where the upper sign corresponds to $N$ and the lower one to $N+1$. The energy of the ground  state is
\bem
\epsilon_N =-\mu' \left(N+{1\over 2}\right) - \sqrt{{\mu'^2\over 4}+z^2J^2(N+1) |\psi|^2},
    \eem{}
where $\mu'=\mu -Un$. The average number of particles per one node is a function of $\mu'$:
\bem
\langle \hat N \rangle=N+{1\over 2} + \frac{\mu'}{ 2 \sqrt{\mu'^2+4z^2J^2(n+1) |\psi|^2} }.
   \eem{}
The self-consistency equation follows either from the minimization of the total energy with respect to $\psi$ or from the condition that $\psi$ is the average value of the operator $\hat b$:
\bem
\psi=\langle \hat b \rangle= \frac{zJ(N+1)}{2  \sqrt{{\mu'^2\over 4}+z^2J^2(N+1) |\psi|^2} }\psi .
   \eem{}
A non-trivial (i.e., non-zero) solution of this equation is
\be
|\psi|^2 ={N+1\over  4}- {\mu'^2\over 4z^2J^2(N+1)}.
   \ee{}

The eigenvalue $\epsilon_N$ of the Hamiltonian (\ref{BH}) determines the  Gibbs thermodynamic potential  $G_N= zJ|\psi|^2 +\epsilon_N$ of the grand canonical ensemble per one node. It is useful to go from the grand canonical ensemble with the Gibbs potential being a function of $\mu$ to   the canonical ensemble where the energy density is a function of the particle number density $n$. Then the energy per node is
\bem
E_N=G_N+\mu N={UN^2\over 2}+UNN_e
\nonumber \\
+zJ\left(|\psi|^2  -  2\sqrt{N+1}\sqrt{\frac{1}{ 4 }-N_e^2} |\psi|\right),
     \eem{En} 
      where $N_e= \langle \hat N \rangle -N-{1\over 2}$.
      The energy has a minimum at
\be
|\psi|^2=(N+1)\left(\frac{1}{ 4 }-N_e^2\right).
   \ee{}
As in any second-order phase transition, $\psi$ vanishes at the phase transition lines, where $N_e=\pm {1\over 2}$ and the number of particles reaches $N$ at the lower border and $N+1$ at the upper one. But in contrast to the Landau-Lifshitz theory of the second order transitions, there is no analytic expansion in $\psi$ near the critical temperature because of the term linear in $|\psi|$.

For the transition to the continuous model, let us consider the effect of slow phase variation from node to node. Assuming that $\psi_i =|\psi|e^{i \bm k \cdot \bm r_i}$ where $\bm r_i$ is the position vector of the $i$th node, one obtains for the square lattice with the number $z_0=4$ of nearest neighbors:
\be
z=2\cos(k_x a)+2\cos(k_y a) \approx 4- k^2 a^2.
    \ee{}
Since the wave vector $\bm k$ corresponds to the gradient operator $\bm \nabla$ in the configurational space one obtains in the continuum limit for small $\bm k$ the  Hamiltonian  (\ref{contHh}) with 
\bem
\tilde n ={2m\over \hbar^2} J (N+1)\left(\frac{1}{ 4 }-n_e^2a^4\right),~~E_c(n)={2\hbar ^2 \tilde n\over m},
        \eem{}
where $n_e=N_e/a^2=n-\left(N+{1\over 2}\right)/a^2$ is the effective density, and constant terms in the energy were ignored.     This allows to find the density dependent factor in the expression (\ref{Mf}) for the Magnus force:
\be
{d\tilde n\over dn} \tilde n=-{8m^2\over \hbar^4} J^2a^4 (N+1)^2n_e\left(\frac{1}{ 4 }-n_e^2a^4\right). 
       \ee{MA}
A remarkable feature of the Magnus force in the beaks of the superfluid phase is that its sign can be inverse with respect to that dictated by the sign of velocity circulation around the vortex. This happens  in the upper halves of the beaks as shown in Fig.~\ref{f12-0}. The regions of the inverse Magnus force neighbor any insulator lobe from below, where $n_e $ is positive. Since at upper borders of the lobes $n_e$ is negative, the line $n_e=0$, where the Magnus force changes its sign, must end somewhere at the border of the lobe. In Fig.~\ref{f12-0} it is shown by a dashed line for the beak between the $N=1$ and $N=2$ lobes. 

\section{Conclusions and discussion}

We derived the transverse (Magnus and Lorentz) forces  on the vortex from the balances of momenta in the continuous approximation for lattice models of superfluids. The two forces are obtained from two different conservation laws, one for the true momentum of particles in the lattice (Magnus force), another for the quasimomentum  (Lorentz force) known from the Bloch band theory. The calculated Magnus force vanishes for the Josephson junction array where the particle-hole symmetry forbids  any Magnus force. The theory was applied for studying vortex dynamics in the Bose-Hubbard model.  In some areas of the phase diagram close to the superfluid--Mott insulator transition the calculated  Magnus force has an inverse sign with respect to the sign dictated by the  velocity circulation around the vortex as has been already revealed earlier. \cite{Auer,Lind}.

Our approach  was based on (i) the continuous approximation for  the lattice model  and on (ii)  the assumption that there is no momentum exchange between the liquid and the lattice if the superfluid is at rest with respect to the lattice. One cannot take validness of these two assumptions for granted. Among most important effects beyond the continuous approximation is intrinsic pinning, which impedes free motion of vortices in the lattice. Therefore one can use our theory if the forces on the vortex are not too weak: they must be higher than the depinning threshold. This may be in conflict with another restriction on the theory that the superfluid velocities are much lower than their critical value. This issue needs a further analysis. Anyway, the theory provides correct results in the two opposite limits: (i) the Galilean invariant liquid with the maximum Magnus force, and (ii) the Josephson junction array with particle-hole symmetry with the zero Magnus force. Therefore, despite possible inaccuracy of assumptions made at its derivation, the theory can serve at least as a reasonable interpolation between these two extreme cases. 

The Magnus force leads to the Hall effect if particles  have a charge $q$. The electric field is determined by the vortex velocity: ${\cal \bm E}={1\over c} [\bm B\times \bm v_L]$.  The value of the Hall conductivity $\sigma_H= j_q /{\cal E}$, where $j_q ={q\over m}j$   is the charge current, depends on the amplitude of the Magnus force. According to our analysis within the Bloch band theory (Sec.~\ref{Bloch}), 
  the Hall conductivity $\sigma_H =(m/m^*)^2qnc /B$ is by the factor $(m/m^*)^2$ less than the Hall conductivity $qnc /B$ known for normal electrons in solids and derived from the same Bloch band theory as used by us. There is no conflict between these two results.
 In the normal electron liquid the magnetic force lines (counterparts of our vortex lines)  move with the same velocity as  charges, i..e., with the group velocity $\bm v_g=\hbar \bm k/m^*$ in the Bloch band, in accordance with Helmholtz's theorem of classic hydrodynamics. In fact, it is the only relevant velocity, since there is no coherent phase, which determines the superfluid velocity $v_s={\hbar \over m}\bm  \nabla \varphi $.  In the superfluid case  Helmholtz's theorem is not valid in general, and the velocity  $\bm v_L$  is a velocity of a phase singularity, which is by the factor $m^*/m$ less than the velocity $\bm v_g$. The uniform magnetic field is crucial for dynamics of normal electrons making them to move along cycloid trajectories.  In dynamics of superconducting vortices the nonuniform magnetic field is localized in fluxons and is commonly neglected, as being  weak compared to  the effect of the phase gradient around the phase singularity \cite{Kop}. 
 
 In the light of this connection between the Magnus force and the Hall conductivity let us  compare our theory with that of  Lindner {\em et al.} \cite{Auer}  and Huber and Lindner \cite{Lind}, who have already  noticed 
 that in the Bose--Hubbard model for charged particles  the Hall conductivity changes its sign together with the sign of $n_e$.\footnote{Note that  Huber and Lindner \cite{Lind} used the name ``Magnus force'' for the force proportional to the superfluid velocity $\bm v_s$ but not to the vortex velocity $\bm v_L$. This disagrees with the nomenclature usually used in the theory superconductivity.  According to this nomenclature used in our paper, the force $\propto \bm v_s$ is the Lorentz force and the  force $\propto \bm v_L$ is the Magnus force (see Sec.~\ref{Intr}).}  However, in their theory the Hall conductivity $\sigma_H$ remains constant at the line $n_e=0$. So the change of the $\sigma_H$ sign is accompanied by a jump of $\sigma_H$, whereas our analysis shows that $\sigma_H $, which is proportional the Magnus force amplitude, must be continuous at $n_e=0$ [see \eq{MA}]. Moreover, our  theory predicts the Hall conductivity, which differs from that  in Refs.~\onlinecite{Auer,Lind}  by the factor $(m/m^*)^2$ proportional to $J^2$ [see \eq{mass}]. The factor can be very small in the tight-binding limit. This is the same factor, which differentiates our Hall conductivity from that of the normal liquid (see the previous paragraph). The Hall conductivity of Refs.~\onlinecite{Auer,Lind} far from the superfluid--insulator transition would be obtained if one considered the effective mass as a true mass of particles in a superfluid  and the quasimomentum as a true momentum. In fact this means that in the theory of Refs.\onlinecite{Auer,Lind} the broken Galilean invariance does not suppress the Magnus force. A possible source of disagreement is that the theory of these papers used topological arguments without directly addressing the momentum balance. 
 
 In the past there were other attempts to derive the Magnus force in lattice superfluids from topology. In particular, the topological origin of the first term $-\hbar n \dot \varphi$ in the Lagrangian (\ref{contL}), which is called the Wess--Zumino term, was widely discussed in the literature \cite{VolB}. The arguments were about whether the total liquid density    $n$ must be replaced by some other density. It is evident that adding any constant $C$ to the density $n$ in the Wess--Zumino term does not affect the Hamilton equations (\ref{nt}) and (\ref{nph}). 
However the role of the Wess--Zumino term changes after transition from the continuous model in terms of fields to the reduced description in terms of the vortex coordinates $\bm r_L(x_L,y_L)$.  This leads to substitution of the phase field $\varphi_v(\bm r-\bm r_L)$ for a  vortex moving with the velocity $\bm v_L=d\bm r_L/dt$ into the Wess--Zumino term  and its integration over the whole space. Bearing in mind that $\dot \varphi_v=-(\bm v_L \cdot \bm \nabla )\varphi_v$, the Wess--Zumino term becomes 
\bem
L_{WZ} =-\hbar (n+C) \bm v_L \cdot [\hat z\times \bm r_L].~~~~
         \eem{}
Varying the total  Lagrangian  of the vortex  with respect to $\bm  r_L(t)$, one obtains the  equation 
of vortex motion with the effective Magnus force $\propto (n+C)$. So the constant $C$ does matter for the value of the Magnus force. It was argued that the contribution $\propto C$ is of topological origin and can be found from the topological analysis \cite{VolB}.
We think that there is no general principle, which dictates the charge in the Wess--Zumino term. It is not the undefined Wess-Zumino term that determines what and whether the transverse force is, but vice versa; one must derive the transverse force from dynamical equations and only after this does one know what Wess-Zumino term should be in the vortex Lagrangian. In  the Galilean and translational invariant liquid one obtains from the momentum-conservation law that the amplitude of the Magnus force is proportional to the total density. Then only the latter enters the Wess--Zumino term and $C=0$. On the other hand, if the Magnus force vanishes, then  the  Wess--Zumino term vanishes also ($C=-n$). 

  The suppression of the Magnus force and the Hall effect because of broken Galilean invariance  in periodic potentials in some sense is similar to the suppression of the Magnus force by the Kopnin--Kravtsov force when Galilean invariance is broken by impurities \cite{Kop}. The existence of the Kopnin--Kravtsov force was also rejected in the past on the basis of some topological analysis connecting the Magnus force with the Berry phase.\cite{AT} Later  it was realized that although this connection definitely exists and is very important a proper calculation  of the Berry phase itself requires in fact knowledge of the Magnus force, which one can obtain only after the dynamical analysis based on the momentum balance. This was demonstrated on the example of the Iordanskii force (the transverse force on the vortex produced by normal quasiparticles) \cite{Th,Magn}, which was also rejected in the original Berry-phase analysis.\cite{AT}

Despite an analogy between  suppression of the Magnus force by a periodic potential discussed in the present paper  and  suppression of the Magnus force by a random potential from impurities in the Kopnin--Kravtsov theory, one should not ignore an important difference between the two cases. The Kopnin--Kravtsov force originated from bound states in vortex cores in Fermi superfluids. In the cases considered in the present paper there were no core bound states, since the Bose--Hubbard model is for Bose superfluids where vortices have no bound states, whereas in the Josephson junction array, which consists of islands of the Fermi superfluid,  vortices have no singular cores. This 
provides a ground for thinking that the existence of core bound states is not critical for suppression of the Magnus force. Therefore, although our analysis addressed ideal strictly periodical lattices, its conclusion about suppression of the Magnus force can  be relevant also both  for Fermi and Bose superfluids in random potentials (e.g., superfluids in porous media or on disordered substates), independently from whether core bound states exist or not.

\begin{acknowledgments}
I thank Ehud Altman, Assa Auerbach, and Netanel Lindner for interesting discussions. The work was supported by the grant of the Israel Academy of Sciences and Humanities and by the FP7 program Microkelvin of the European Union.
\end{acknowledgments}


\begin{thebibliography}{17}%
\makeatletter
\providecommand \@ifxundefined [1]{%
 \@ifx{#1\undefined}
}%
\providecommand \@ifnum [1]{%
 \ifnum #1\expandafter \@firstoftwo
 \else \expandafter \@secondoftwo
 \fi
}%
\providecommand \@ifx [1]{%
 \ifx #1\expandafter \@firstoftwo
 \else \expandafter \@secondoftwo
 \fi
}%
\providecommand \natexlab [1]{#1}%
\providecommand \enquote  [1]{``#1''}%
\providecommand \bibnamefont  [1]{#1}%
\providecommand \bibfnamefont [1]{#1}%
\providecommand \citenamefont [1]{#1}%
\providecommand \href@noop [0]{\@secondoftwo}%
\providecommand \href [0]{\begingroup \@sanitize@url \@href}%
\providecommand \@href[1]{\@@startlink{#1}\@@href}%
\providecommand \@@href[1]{\endgroup#1\@@endlink}%
\providecommand \@sanitize@url [0]{\catcode `\\12\catcode `\$12\catcode
  `\&12\catcode `\#12\catcode `\^12\catcode `\_12\catcode `\%12\relax}%
\providecommand \@@startlink[1]{}%
\providecommand \@@endlink[0]{}%
\providecommand \url  [0]{\begingroup\@sanitize@url \@url }%
\providecommand \@url [1]{\endgroup\@href {#1}{\urlprefix }}%
\providecommand \urlprefix  [0]{URL }%
\providecommand \Eprint [0]{\href }%
\providecommand \doibase [0]{http://dx.doi.org/}%
\providecommand \selectlanguage [0]{\@gobble}%
\providecommand \bibinfo  [0]{\@secondoftwo}%
\providecommand \bibfield  [0]{\@secondoftwo}%
\providecommand \translation [1]{[#1]}%
\providecommand \BibitemOpen [0]{}%
\providecommand \bibitemStop [0]{}%
\providecommand \bibitemNoStop [0]{.\EOS\space}%
\providecommand \EOS [0]{\spacefactor3000\relax}%
\providecommand \BibitemShut  [1]{\csname bibitem#1\endcsname}%
\let\auto@bib@innerbib\@empty
\bibitem [{\citenamefont {Sonin}(1987)}]{RMP}%
  \BibitemOpen
  \bibfield  {author} {\bibinfo {author} {\bibfnamefont {E.~B.}\ \bibnamefont
  {Sonin}},\ }\href@noop {} {\bibfield  {journal} {\bibinfo  {journal} {Rev.
  Mod. Phys.}\ }\textbf {\bibinfo {volume} {59}},\ \bibinfo {pages} {87}
  (\bibinfo {year} {1987})}\BibitemShut {NoStop}%
\bibitem [{\citenamefont {Kopnin}(2001)}]{Kop}%
  \BibitemOpen
  \bibfield  {author} {\bibinfo {author} {\bibfnamefont {N.~B.}\ \bibnamefont
  {Kopnin}},\ }\href@noop {} {\emph {\bibinfo {title} {Theory of nonequilibrium
  superconductivity}}}\ (\bibinfo  {publisher} {Oxford University Press},\
  \bibinfo {year} {2001})\BibitemShut {NoStop}%
\bibitem [{\citenamefont {Sonin}(2002)}]{Magn}%
  \BibitemOpen
  \bibfield  {author} {\bibinfo {author} {\bibfnamefont {E.~B.}\ \bibnamefont
  {Sonin}},\ }in\ \href@noop {} {\emph {\bibinfo {booktitle} {Vortices in
  Unconventional Superconductors and Superfluids}}},\ \bibinfo {editor} {edited
  by\ \bibinfo {editor} {\bibfnamefont {R.~P.}\ \bibnamefont {Huebener}},
  \bibinfo {editor} {\bibfnamefont {N.}~\bibnamefont {Schopohl}}, \ and\
  \bibinfo {editor} {\bibfnamefont {G.~E.}\ \bibnamefont {Volovik}}}\ (\bibinfo
   {publisher} {Springer-Verlag},\ \bibinfo {year} {2002})\ pp.\ \bibinfo
  {pages} {119--145}\BibitemShut {NoStop}%
\bibitem [{\citenamefont {Volovik}(2003)}]{VolB}%
  \BibitemOpen
  \bibfield  {author} {\bibinfo {author} {\bibfnamefont {G.~E.}\ \bibnamefont
  {Volovik}},\ }\href@noop {} {\emph {\bibinfo {title} {The universe in a
  Helium droplet}}}\ (\bibinfo  {publisher} {Oxford University Press},\
  \bibinfo {year} {2003})\BibitemShut {NoStop}%
\bibitem [{\citenamefont {Eckern}\ and\ \citenamefont {Schmid}(1989)}]{Eck}%
  \BibitemOpen
  \bibfield  {author} {\bibinfo {author} {\bibfnamefont {U.}~\bibnamefont
  {Eckern}}\ and\ \bibinfo {author} {\bibfnamefont {A.}~\bibnamefont
  {Schmid}},\ }\href@noop {} {\bibfield  {journal} {\bibinfo  {journal} {Phys.
  Rev. B}\ }\textbf {\bibinfo {volume} {39}},\ \bibinfo {pages} {6441}
  (\bibinfo {year} {1989})}\BibitemShut {NoStop}%
\bibitem [{\citenamefont {van~der Zant}\ \emph {et~al.}(1992)\citenamefont
  {van~der Zant}, \citenamefont {Fritschy}, \citenamefont {Orlando},\ and\
  \citenamefont {Mooij}}]{2}%
  \BibitemOpen
  \bibfield  {author} {\bibinfo {author} {\bibfnamefont {H.~S.~J.}\
  \bibnamefont {van~der Zant}}, \bibinfo {author} {\bibfnamefont {F.~C.}\
  \bibnamefont {Fritschy}}, \bibinfo {author} {\bibfnamefont {T.~P.}\
  \bibnamefont {Orlando}}, \ and\ \bibinfo {author} {\bibfnamefont {J.~E.}\
  \bibnamefont {Mooij}},\ }\href@noop {} {\bibfield  {journal} {\bibinfo
  {journal} {Europhys. Lett.}\ }\textbf {\bibinfo {volume} {18}},\ \bibinfo
  {pages} {343} (\bibinfo {year} {1992})}\BibitemShut {NoStop}%
\bibitem [{\citenamefont {Sonin}(1997)}]{PRB7}%
  \BibitemOpen
  \bibfield  {author} {\bibinfo {author} {\bibfnamefont {E.~B.}\ \bibnamefont
  {Sonin}},\ }\href@noop {} {\bibfield  {journal} {\bibinfo  {journal} {Phys.
  Rev. B}\ }\textbf {\bibinfo {volume} {55}},\ \bibinfo {pages} {485} (\bibinfo
  {year} {1997})}\BibitemShut {NoStop}%
\bibitem [{\citenamefont {Fisher}\ \emph {et~al.}(1989)\citenamefont {Fisher},
  \citenamefont {Weichman}, \citenamefont {Grinstein},\ and\ \citenamefont
  {Fisher}}]{FishH}%
  \BibitemOpen
  \bibfield  {author} {\bibinfo {author} {\bibfnamefont {M.~P.~A.}\
  \bibnamefont {Fisher}}, \bibinfo {author} {\bibfnamefont {P.~B.}\
  \bibnamefont {Weichman}}, \bibinfo {author} {\bibfnamefont {G.}~\bibnamefont
  {Grinstein}}, \ and\ \bibinfo {author} {\bibfnamefont {D.~S.}\ \bibnamefont
  {Fisher}},\ }\href {\doibase 10.1103/PhysRevB.40.546} {\bibfield  {journal}
  {\bibinfo  {journal} {Phys. Rev. B}\ }\textbf {\bibinfo {volume} {40}},\
  \bibinfo {pages} {546} (\bibinfo {year} {1989})}\BibitemShut {NoStop}%
\bibitem [{\citenamefont {Ueda}(2010)}]{Ued}%
  \BibitemOpen
  \bibfield  {author} {\bibinfo {author} {\bibfnamefont {M.}~\bibnamefont
  {Ueda}},\ }\href@noop {} {\emph {\bibinfo {title} {Fundamentals and new
  frontiers of {Bose--Einstein} condensation}}}\ (\bibinfo  {publisher} {World
  {Sciientific}},\ \bibinfo {year} {2010})\BibitemShut {NoStop}%
\bibitem [{\citenamefont {Lindner}\ \emph {et~al.}(2010)\citenamefont
  {Lindner}, \citenamefont {Auerbach},\ and\ \citenamefont {Arovas}}]{Auer}%
  \BibitemOpen
  \bibfield  {author} {\bibinfo {author} {\bibfnamefont {N.}~\bibnamefont
  {Lindner}}, \bibinfo {author} {\bibfnamefont {A.}~\bibnamefont {Auerbach}}, \
  and\ \bibinfo {author} {\bibfnamefont {D.~P.}\ \bibnamefont {Arovas}},\
  }\href {\doibase 10.1103/PhysRevB.82.134510} {\bibfield  {journal} {\bibinfo
  {journal} {Phys. Rev. B}\ }\textbf {\bibinfo {volume} {82}},\ \bibinfo
  {pages} {134510} (\bibinfo {year} {2010})}\BibitemShut {NoStop}%
\bibitem [{\citenamefont {Huber}\ and\ \citenamefont {Lindner}(2011)}]{Lind}%
  \BibitemOpen
  \bibfield  {author} {\bibinfo {author} {\bibfnamefont {S.~D.}\ \bibnamefont
  {Huber}}\ and\ \bibinfo {author} {\bibfnamefont {N.~H.}\ \bibnamefont
  {Lindner}},\ }\href@noop {} {\bibfield  {journal} {\bibinfo  {journal}
  {PNAS}\ }\textbf {\bibinfo {volume} {108}},\ \bibinfo {pages} {19925}
  (\bibinfo {year} {2011})}\BibitemShut {NoStop}%
\bibitem [{\citenamefont {Ilisavskii}\ \emph {et~al.}(2001)\citenamefont
  {Ilisavskii}, \citenamefont {Goltsev}, \citenamefont {Dyakonov},
  \citenamefont {Popov}, \citenamefont {Yakhkind}, \citenamefont {Dyakonov},
  \citenamefont {Gier\l{}owski}, \citenamefont {Klimov}, \citenamefont
  {Lewandowski},\ and\ \citenamefont {Szymczak}}]{AEE}%
  \BibitemOpen
  \bibfield  {author} {\bibinfo {author} {\bibfnamefont {Y.}~\bibnamefont
  {Ilisavskii}}, \bibinfo {author} {\bibfnamefont {A.}~\bibnamefont {Goltsev}},
  \bibinfo {author} {\bibfnamefont {K.}~\bibnamefont {Dyakonov}}, \bibinfo
  {author} {\bibfnamefont {V.}~\bibnamefont {Popov}}, \bibinfo {author}
  {\bibfnamefont {E.}~\bibnamefont {Yakhkind}}, \bibinfo {author}
  {\bibfnamefont {V.~P.}\ \bibnamefont {Dyakonov}}, \bibinfo {author}
  {\bibfnamefont {P.}~\bibnamefont {Gier\l{}owski}}, \bibinfo {author}
  {\bibfnamefont {A.}~\bibnamefont {Klimov}}, \bibinfo {author} {\bibfnamefont
  {S.~J.}\ \bibnamefont {Lewandowski}}, \ and\ \bibinfo {author} {\bibfnamefont
  {H.}~\bibnamefont {Szymczak}},\ }\href {\doibase
  10.1103/PhysRevLett.87.146602} {\bibfield  {journal} {\bibinfo  {journal}
  {Phys. Rev. Lett.}\ }\textbf {\bibinfo {volume} {87}},\ \bibinfo {pages}
  {146602} (\bibinfo {year} {2001})}\BibitemShut {NoStop}%
\bibitem [{\citenamefont {Cerda-M\`endez}\ \emph {et~al.}(2010)\citenamefont
  {Cerda-M\`endez}, \citenamefont {Krizhanovskii}, \citenamefont {Wouters},
  \citenamefont {Bradley}, \citenamefont {Biermann}, \citenamefont {Guda},
  \citenamefont {Hey}, \citenamefont {Santos}, \citenamefont {Sarkar},\ and\
  \citenamefont {Skolnick}}]{AcusEx}%
  \BibitemOpen
  \bibfield  {author} {\bibinfo {author} {\bibfnamefont {E.~A.}\ \bibnamefont
  {Cerda-M\`endez}}, \bibinfo {author} {\bibfnamefont {D.~N.}\ \bibnamefont
  {Krizhanovskii}}, \bibinfo {author} {\bibfnamefont {M.}~\bibnamefont
  {Wouters}}, \bibinfo {author} {\bibfnamefont {R.}~\bibnamefont {Bradley}},
  \bibinfo {author} {\bibfnamefont {K.}~\bibnamefont {Biermann}}, \bibinfo
  {author} {\bibfnamefont {K.}~\bibnamefont {Guda}}, \bibinfo {author}
  {\bibfnamefont {R.}~\bibnamefont {Hey}}, \bibinfo {author} {\bibfnamefont
  {P.~V.}~\bibnamefont {Santos}}, \bibinfo {author} {\bibfnamefont
  {D.}~\bibnamefont {Sarkar}}, \ and\ \bibinfo {author} {\bibfnamefont {M.~S.}\
  \bibnamefont {Skolnick}},\ }\href@noop {} {\bibfield  {journal} {\bibinfo
  {journal} {Phys. Rev. Lett.}\ }\textbf {\bibinfo {volume} {105}},\ \bibinfo
  {pages} {116402} (\bibinfo {year} {2010})}\BibitemShut {NoStop}%
\bibitem [{Note1()}]{Note1}%
  \BibitemOpen
  \bibinfo {note} {We assume that there is the same number of particles at all
  nodes and write $N_i$ without the subscript $i$.}\BibitemShut {Stop}%
\bibitem [{Note2()}]{Note2}%
  \BibitemOpen
  \bibinfo {note} {Note that Huber and Lindner \cite {Lind} used the name
  ``Magnus force'' for the force proportional to the superfluid velocity
  $\protect \bm {v}_s$ but not to the vortex velocity $\protect \bm {v}_L$.
  This disagrees with the nomenclature usually used in the theory
  superconductivity. According to this nomenclature used in our paper, the
  force $\propto \protect \bm {v}_s$ is the Lorentz force and the force
  $\propto \protect \bm {v}_L$ is the Magnus force (see Sec.~\ref
  {Intr}).}\BibitemShut {Stop}%
\bibitem [{\citenamefont {Ao}\ and\ \citenamefont {Thouless}(1993)}]{AT}%
  \BibitemOpen
  \bibfield  {author} {\bibinfo {author} {\bibfnamefont {P.}~\bibnamefont
  {Ao}}\ and\ \bibinfo {author} {\bibfnamefont {D.~J.}\ \bibnamefont
  {Thouless}},\ }\href@noop {} {\bibfield  {journal} {\bibinfo  {journal}
  {Phys. Rev. Lett.}\ }\textbf {\bibinfo {volume} {70}},\ \bibinfo {pages}
  {2158} (\bibinfo {year} {1993})}\BibitemShut {NoStop}%
\bibitem [{\citenamefont {Thouless}\ \emph {et~al.}(2001)\citenamefont
  {Thouless}, \citenamefont {Geller}, \citenamefont {Vinen}, \citenamefont
  {Fortin},\ and\ \citenamefont {Rhee}}]{Th}%
  \BibitemOpen
  \bibfield  {author} {\bibinfo {author} {\bibfnamefont {D.~J.}\ \bibnamefont
  {Thouless}}, \bibinfo {author} {\bibfnamefont {M.~R.}\ \bibnamefont
  {Geller}}, \bibinfo {author} {\bibfnamefont {W.~F.}\ \bibnamefont {Vinen}},
  \bibinfo {author} {\bibfnamefont {J.-Y.}\ \bibnamefont {Fortin}}, \ and\
  \bibinfo {author} {\bibfnamefont {S.~W.}\ \bibnamefont {Rhee}},\ }\href@noop
  {} {\bibfield  {journal} {\bibinfo  {journal} {Phys. Rev. B}\ }\textbf
  {\bibinfo {volume} {63}},\ \bibinfo {pages} {224504} (\bibinfo {year}
  {2001})}\BibitemShut {NoStop}%
\end{thebibliography}

%

\end{document}